\documentclass[acmsmall,screen]{acmart}
\usepackage[shortlabels]{enumitem}
\usepackage{fancybox, graphicx}
\usepackage{color}
\usepackage{listings}
\usepackage{tcolorbox}
\usepackage{balance}
\usepackage{tabularx}
\usepackage{multirow}
\usepackage{xspace}
\usepackage{tikz}
\usepackage{textcomp}
\usepackage{booktabs}
\usetikzlibrary{patterns}
\usepackage{url}

\definecolor{Gray}{gray}{0.9}
\definecolor{OliveGreen}{cmyk}{0.64,0,0.95,0.40}
\definecolor{lightlightgray}{gray}{0.9}
\usepackage[scaled=0.85]{beramono}
\definecolor{javared}{rgb}{0.6,0,0} 
\definecolor{javagreen}{rgb}{0.25,0.5,0.35} 
\definecolor{javapurple}{rgb}{0.5,0,0.2} 
\definecolor{javadocblue}{rgb}{0.25,0.35,0.75} 

\lstdefinelanguage{java-pretty}
{
  language=java,
  numbers=left,
  frame=shadowbox,
  rulesepcolor= \color{red!20!green!20!blue!20},
  basicstyle=\footnotesize\ttfamily,
  numberstyle=\scriptsize,
  breaklines=true,
  columns=fullflexible,
  xleftmargin=16pt,
  showstringspaces=false,
  keywordstyle=\color{blue}\bfseries,
  stringstyle=\color{javared},
  commentstyle=\color{javagreen},
  morecomment=[s][\color{javadocblue}]{/**}{*/},
}
\colorlet{punct}{red!60!black}
\definecolor{background}{HTML}{EEEEEE}
\definecolor{delim}{RGB}{20,105,176}
\colorlet{numb}{magenta!60!black}
\lstdefinelanguage{json}{
    basicstyle=\normalfont\ttfamily,
    numbers=left,
    numberstyle=\scriptsize,
    stepnumber=1,
    numbersep=8pt,
    showstringspaces=false,
    breaklines=true,
    frame=lines,
    literate=
     *{:}{{{\color{punct}{:}}}}{1}
      {,}{{{\color{punct}{,}}}}{1}
      {\{}{{{\color{delim}{\{}}}}{1}
      {\}}{{{\color{delim}{\}}}}}{1}
      {[}{{{\color{delim}{[}}}}{1}
      {]}{{{\color{delim}{]}}}}{1},
}

\lstdefinelanguage{json-pretty}
{
  language=json,
  numbers=left,
  frame=shadowbox,
  rulesepcolor= \color{red!20!green!20!blue!20},
  basicstyle=\footnotesize\ttfamily,
  numberstyle=\scriptsize,
  breaklines=true,
  columns=fullflexible,
  xleftmargin=16pt,
  showstringspaces=false,
  keywordstyle=\color{blue}\bfseries,
  stringstyle=\color{javared},
  commentstyle=\color{javagreen},
  morecomment=[s][\color{javadocblue}]{/**}{*/},
}

\lstdefinelanguage{diff}{
    basicstyle=\ttfamily\small,
    morecomment=[f][\color{diffstart}]{@@},
    morecomment=[f][\color{javagreen}]{+\ },
    morecomment=[f][\color{javared}]{-\ },
  }

\newcommand{\InputWithSpace}[1]{\bgroup\def\arraystretch{1.15}\input{#1}\egroup}

\setcopyright{acmcopyright}
\copyrightyear{2023}
\acmYear{2023}
\acmDOI{XXXXXXX.XXXXXXX}

\acmJournal{JACM}
\acmVolume{37}
\acmNumber{4}
\acmArticle{1}
\acmMonth{8}

\begin{document}

\title{Evaluating Pre-trained Language Models for Repairing API Misuses}

\author{Ting Zhang}
\affiliation{%
  \institution{Singapore Management University}
  \country{Singapore}
}
\email{tingzhang.2019@phdcs.smu.edu.sg}
\orcid{0000-0002-6001-1372}

\author{Ivana Clairine Irsan}
\affiliation{%
    \institution{Singapore Management University}
  \country{Singapore}
}
\email{ivanairsan@smu.edu.sg}
\orcid{0000-0001-6350-2700}

\author{Ferdian Thung}
\affiliation{%
    \institution{Singapore Management University}
  \country{Singapore}
}
\email{ferdianthung@smu.edu.sg}
\orcid{0000-0002-5566-3819}

\author{David Lo}
\affiliation{%
  \institution{Singapore Management University}
  \country{Singapore}
}
\email{davidlo@smu.edu.sg}
\orcid{0000-0002-4367-7201}

\author{Asankhaya Sharma}
\affiliation{%
  \institution{Singapore Management University}
  \country{Singapore}
}
\email{asankhayas@smu.edu.sg}
\orcid{0000-0001-6185-0410}

\author{Lingxiao Jiang}
\affiliation{%
  \institution{Singapore Management University}
  \country{Singapore}
}
\email{lxjiang@smu.edu.sg}
\orcid{0000-0002-4336-8548}

\renewcommand{\shortauthors}{Zhang et al.}




\begin{abstract}
API misuses often lead to software bugs, crashes, and vulnerabilities. While several API misuse detectors have been proposed, there are no automatic repair tools specifically designed for this purpose. In a recent study, test-suite-based automatic program repair (APR) tools were found to be ineffective in repairing API misuses. Still, since the study focused on non-learning-aided APR tools, it remains unknown whether learning-aided APR tools are capable of fixing API misuses. In recent years, pre-trained language models (PLMs) have succeeded greatly in many natural language processing tasks. There is a rising interest in applying PLMs to APR.  However, there has not been any study that investigates the effectiveness of PLMs in repairing API misuse. 

To fill this gap, we conduct a comprehensive empirical study on 11 learning-aided APR tools, which include 9 of the state-of-the-art general-purpose PLMs and two APR tools. We evaluate these models with an API-misuse repair dataset, consisting of two variants. Our results show that PLMs perform better than the studied APR tools in repairing API misuses. Among the 9 pre-trained models tested, CodeT5 is the best performer in the exact match. We also offer insights and potential exploration directions for future research.
\end{abstract}

\maketitle

\newcommand{\XSpace}[1]{}

\makeatletter
\newenvironment{btHighlight}[1][]
{\begingroup\tikzset{bt@Highlight@par/.style={#1}}\begin{lrbox}{\@tempboxa}}
{\end{lrbox}\bt@HL@box[bt@Highlight@par]{\@tempboxa}\endgroup}

\newcommand\btHL[1][]{%
  \begin{btHighlight}[#1]\bgroup\aftergroup\bt@HL@endenv%
}
\def\bt@HL@endenv{%
  \end{btHighlight}%
  \egroup
}
\newcommand{\bt@HL@box}[2][]{%
  \tikz[#1]{%
    \pgfpathrectangle{\pgfpoint{1pt}{0pt}}{\pgfpoint{\wd #2}{\ht #2}}%
    \pgfusepath{use as bounding box}%
    \node[anchor=base west, fill=yellow!30,outer sep=0pt,inner xsep=1pt, inner ysep=0pt, rounded corners=0pt, minimum height=\ht\strutbox+1pt,#1]{\raisebox{1pt}{\strut}\strut\usebox{#2}};
  }%
}
\makeatother
\lstdefinestyle{Highlight}{
    moredelim=**[is][\btHL]{`}{`},
    moredelim=**[is][{\btHL[fill=orange!50]}]{´}{´},
    moredelim=**[is][{\btHL[fill=red!50]}]{@}{@},
}

\newcommand{\wholedata}{{\textit{complete data}}}
\newcommand{\singledata}{{\textit{single-line data}}}

\section{Introduction}
\label{sec:introduction}

Application Programming Interfaces (APIs) are widely used in many types of software systems, such as web applications~\cite{bae2014safewapi} and mobile applications~\cite{haryono2020automatic}.
An API misuse refers to the use of an API that violates explicit or implicit \textit{usage constraints} of the API~\cite{amann2016mubench,amann2018systematic,kechagia2021evaluating}.
API misuses are a prevalent issue in software development, as it is not always easy to use APIs correctly.
Misuse of APIs can lead to software bugs, crashes, and security vulnerabilities~\cite{gu2019empirical,yun2016apisan}.
For instance, misuse of APIs of Secure Sockets Layer implementations (such as JSSE, OpenSSL, and GnuTLS) can lead to man-in-the-middle attacks~\cite{georgiev2012most}.
According to a recent study~\cite{gu2019empirical}, 17\% of the bug-fixing commits are related to API misuses.
To alleviate this problem, in recent years, many API misuse detection approaches have been proposed~\cite{yun2016apisan,li2021arbitrar,kechagia2019effective}.
Despite these recent efforts, API misuses remain prevalent~\cite{legunsen2016good,zhang2018code}.

While several approaches have been proposed to detect API misuse automatically~\cite{yun2016apisan,li2021arbitrar,kechagia2019effective}, substantial extra effort is needed to fix API misuses manually.
It is desirable to have an automatic approach that can help developers repair API misuses.
Unfortunately, despite its importance, there is a lack of research in developing specialized techniques for repairing API misuse.
Kechagia et al.~\cite{kechagia2021evaluating} made the most recent and relevant endeavor: they conducted an empirical study on 14 Java test-suite-based APR techniques regarding their ability to repair API misuses.
Test-suite-based APR techniques take an input of the buggy program and a test suite where at least one failing test case is related to the bug and then generate a patch that passes the test suite~\cite{xiong2018identifying}.
They found that, while most of the patches generated by APR techniques are plausible, only a few of them are semantically correct when compared to the patches written by developers.
However, their evaluation excluded \textit{learning-aided} APR techniques, which require extra data for learning fixing patterns.
Thus, how learning-aided APR techniques perform in repairing API misuses remains unknown.

Learning-aided APR usually considers the bug-fixing task as a neural machine translation (NMT) problem and attempts to tackle it by Sequence-to-Sequence (Seq2Seq) learning~\cite{zhang2023survey}.
These techniques attempt to learn from historical fixes to repair bugs.
As the state-of-the-art Seq2Seq architecture, the Transformer model~\cite{vaswani2017attention} has received much attention in software engineering lately, particularly \textit{general-purpose} pre-trained language models of code (PLMs) such as CodeBERT~\cite{codebert}, CodeGPT~\cite{codegpt}, and CodeT5~\cite{codet5}.
These models have been extensively applied to solve various code understanding and generation tasks, including defect detection, code summarization, and code translation~\cite{wang2022no,zeng2022extensive}.
They have also effectively repaired general program errors~\cite{xia2022less,yuan2022circle}.
In addition, with increased access to historical fixes, \textit{specialized} learning-aided APR techniques~\cite{le2016history,long2016automatic} are becoming more popular.
The last couple of years have witnessed the boost of learning-aided-based APR techniques, such as SequenceR~\cite{sequencer}, CoCoNut~\cite{coconut}, and CURE~\cite{cure}.
However, no attempt has been made to apply either general-purpose PLMs or specialized learning-aided APR techniques to repair API misuse bugs.
Note that we consider both general-purpose PLMs and specialized learning-aided APR techniques as learning-aided APR techniques in this paper.

Moreover, the absence of benchmarks that are specifically designed for repairing API misuse bugs hinders research progress.
The newly proposed APR techniques are usually evaluated on general program error benchmarks, such as Defects4J~\cite{just2014defects4j}, QuixBugs~\cite{lin2017quixbugs} and BEARS~\cite{madeiral2019bears}.
Although there are some API misuse bugs in these benchmarks~\cite{kechagia2021evaluating}, they are not the main focus of these benchmarks.
Thus, it lacks emphasis on the evaluation and discussion of repairing API misuse bugs.
Recently, Kechagia et al.~\cite{kechagia2021evaluating} shared a test-suite-based API misuse repair dataset (\textsc{APIRepBench}), which contains 101 API misuses.
However, learning-aided APR techniques typically require a substantial amount of historical fixes for training.
Thus, the limited number of API misuses in \textsc{APIRepBench} is insufficient to train and evaluate the performance of learning-aided APR techniques.
Recently, Li et al.~\cite{li2021large} analyzed API misuses in the wild by extracting API misuses based on 528,546 historical bug-fixing commits from GitHub (from 2011 to 2018).
These commits make it possible to evaluate the performance of learning-aided APR techniques to their full potential.

In this work, we derive an API misuse repair dataset from the dataset introduced by Li et al.~\cite{li2021large}, by removing noises in the original dataset and extracting the buggy and fixed method pairs. 
The dataset contains 118,490 such method pairs. 
We also prepare a subset of the dataset that contains only one-line bugs and call it \singledata. 
We call the full dataset as \textit{complete data}.
We then evaluate the performance of learning-aided APR techniques on this dataset.
To examine whether state-of-the-art PLMs are capable of fixing API misuses in methods, we compare the performance of 9 PLMs on the \textit{complete data} (\textbf{RQ1}).
We also investigate how PLMs and specific APR techniques perform in fixing API misuses in methods with single-line changes, i.e., on the \textit{single-line data} (\textbf{RQ2}).

To the best of our knowledge, we are the first to examine the performance of learning-aided APR techniques on API misuse repair.
Our results show that PLMs are more capable of repairing API misuse than the studied APR techniques.
Among the 9 PLMs tested, CodeT5 was the best performer.
On the \textit{complete data}, CodeT5 achieves an exact match ratio of 3.01.
On the \singledata, CodeT5 achieves an exact match ratio of 8.86. 

Our contributions can be summarized as follows:
\begin{itemize}[leftmargin=*, topsep=1pt, itemsep=1pt]
    \item{We provide a benchmark, which contains 118,490 method pairs on \textit{complete data} and 54,510 method pairs on \singledata. 
    For each method pair, the first method contains API misuse(s), and the second contains the corrected API misuse(s).}
    \item{We evaluated a total of 11 learning-aided program repair techniques on the benchmark.}
    \item{We provide implications and insights on the learning-aided APR techniques' ability to repair API misuse.}
\end{itemize}

The rest of the paper is organized as follows.
We outline the background in Section~\ref{sec:background}.
We elaborate on the details of the experimental setting in Section~\ref{sec:setting} and results in Section~\ref{sec:result}.
We discuss the implications of our work and threats to validity in Section~\ref{sec:discussion}.
In Section~\ref{sec:related}, we list the related works.
We finally conclude our work and list the potential future work in Section~\ref{sec:conclusion}.
\section{Background}
\label{sec:background}

\subsection{Automatic Program Repair}
Automatic Program Repair (APR) aims to fix buggy programs with less manual effort.
It mainly consists of two steps: (1) conducting fault localization to detect the bug and (2) generating the bug fixes.
They are several ways to group the current APR techniques.
We follow the prior works~\cite{kechagia2021evaluating,gao2022program} to categorize APR techniques into three groups: (1) \textit{heuristic-based repair} techniques, (2) \textit{constraint-based repair} techniques, and (3) \textit{learning-aided repair} techniques.

\textit{Heuristic-based repair} technique is also known as \textit{generate-and-validate repair} technique~\cite{wen2018context}.
In this category, APR is treated as a search problem.
The approaches apply heuristic strategies to explore the search space and validate the candidate patches exhaustively.
For example, GenProg~\cite{le2011genprog} randomly selects a set of candidate patches as the initial population and then applies genetic programming to generate patches.
In the patch validation stage, each patch will be validated against a test suite to compute the fitness.
RSRepair~\cite{qi2014strength} replaces the genetic programming in GenProg with a random search.
This category also includes \textit{template-based repair} techniques since both types work similarly~\cite{liu2020efficiency}.
For example, TBar~\cite{liu2019tbar} combines various fix patterns collected from previous studies.
After a fix pattern is selected to repair a bug, a patch is generated and validated by a test suite.
Heuristic-based repair techniques are usually restricted by the pre-defined heuristics, which may either be insufficient to cover all the possible patches or too exhaustive to be efficiently explored~\cite{gao2022program}.

\textit{Constraint-based repair} techniques consider APR as a constraint-solving problem.
Specifically, they formulate the requirement to pass all the test cases as a set of constraints and then solve them to generate the patches~\cite{nguyen2013semfix,xuan2016nopol,mechtaev2016angelix}.
For example, SemFix~\cite{nguyen2013semfix} infers constraints by performing symbolic executions on the supplied test cases.
It then looks for concrete expressions that enable the program to pass the test cases by Satisfiability Modulo Theories (SMT) solvers.
Nopol~\cite{xuan2016nopol} uses angelic debugging to identify conditions in buggy Java programs that, if changed, could allow the program to pass the test suite. 
It then employs an SMT solver to synthesize repairs for these conditions~\cite {long2016automatic}.
Constraint-based repair techniques suffer from the scalability problem since they usually need to conduct a heavy symbolic execution to collect constraints, and solving these constraints can be time-consuming~\cite{gao2022program}.

In this work, we focus on the last category, i.e., \textit{learning-aided repair} techniques, which leverages the previously generated bug fixes.
Usually, a machine-learning or deep-learning model would be utilized to learn the bug-fixing patterns from large corpora of code~\cite{tufano2019empirical}.
Existing learning-aided repair techniques generally treat APR as an NMT problem, which translates the buggy code into the bug fixes~\cite{tufano2019empirical,coconut,sequencer,cure}.
However, the granularity of the bug fixes is different among approaches.
Some approaches work on the method level.
For instance, Tufano et al.~\cite{tufano2019empirical} evaluate an Encoder-Decoder model using the buggy method as the input and the fixed method as the output.
In addition, most PLMs that have been evaluated on program repair also use the same dataset~\cite{codet5,plbart}.
Other approaches work on the line-level~\cite{cure,li2022dear,coconut,recoder,yuan2022circle}.
CoCoNut~\cite{coconut} takes an input of buggy lines and the buggy method as the context, and it outputs the fixed lines.
We will describe more about the techniques that we evaluate in our study in Section~\ref{sec:spec-tools}.

\subsection{API Misuses}
An API \textit{usage} can be categorized into two types, i.e., directly calling API methods or instantiating objects from API classes~\cite{nguyen2010graph}.
API \textit{misuses} are as violations of (implicit) usage constraints of APIs~\cite{amann2018systematic}.
Recent years have witnessed a number of API-misuse detectors~\cite{yun2016apisan,li2021arbitrar,kechagia2019effective,kang2021active}.
Generally speaking, there are three types of API misuse detection approaches, i.e., static detectors, which detect API misuses by statically analyzing the source code or binary code~\cite{monperrus2013detecting}; dynamic detectors, which instead detect API misuses by dynamic analysis~\cite{pradel2012leveraging}; and the third one combines mining with the static detector~\cite{pradel2012statically}.

Regardless of the program analysis technique, existing techniques either (1) require API specification: They consider the violation of a specification as an API misuse. 
The drawback is that the specification is usually incomplete and hard to obtain~\cite{li2021arbitrar}; 
or (2) do not require API specification:
Given a large-scale code corpus, they consider the majority usage pattern valid.
The use of an API is considered to be a misuse if it does not follow the majority usage pattern.
The drawback is that this assumption does not always hold.
For instance, some APIs are rarely used, and the majority usage pattern does not represent valid usage~\cite{kang2021active}.

\subsection{API Misuse Benchmark}
Recently, several API-misuse benchmarks have been proposed.
Amann et al.~\cite{amann2018systematic} proposed the first benchmark named MuBench for API misuse detection.
MuBench encompasses 89 API misuses.
Among them, 77 API misuses are from real-world projects, and the remaining 12 misuses are from the survey they conducted.
The 77 API misuses are collected from (1) existing bug datasets, i.e., BugClassify~\cite{herzig2013s}, Defects4J~\cite{just2014defects4j}, and QACrashFix~\cite{gao2015fixing}; (2) bug-fixing changes from projects on SourceForge and GitHub for misuses of Java Cryptography Extension (JCE) APIs.
MuBench is the first benchmark that has been used to evaluate API-misuse detectors.

\textsc{APIRepBench}~\cite{kechagia2021evaluating} is the first benchmark that can be used for API misuse repair.
It contains 101 API misuses from 29 Java projects.
This benchmark is derived from three existing bug benchmarks, i.e., MuBench~\cite{amann2016mubench}, BEARS~\cite{madeiral2019bears}, and Bugs.jar~\cite{saha2018bugs}.
It only contains API misuses that belong to the \textit{missing} category, such as missing method calls, missing null checks, and missing exception handling.
Due to the limited size of this benchmark, it is insufficient to train and evaluate the learning-aided APR techniques.

More recently, Li et al.~\cite{li2021large} created a large-scale API misuse repair benchmark, which contains 528,546 bug-fixing commits of Java projects from 2011 to 2018.
They extracted fine-grained edit operations on the Abstract Syntax Tree (AST) of the source code.
They also classify the API misuses into different categories, including \textit{missing}, \textit{redundant} and \textit{replaced}.
Based on the categories, they extracted frequent API misuse patterns. 
Since this dataset is the largest and most recent, we derive a dataset that contains method pairs from it. 
We describe the detail in Section~\ref{sec:benchmark}.
\section{Experimental Setup}
\label{sec:setting}

\subsection{Research Questions}
In this study, our primary objective is to investigate the effectiveness of learning-aided repair methods in repairing API misuses, particularly when employing general-purpose PLMs.

With this goal in mind, we aim to answer the following two Research Questions (RQs).

\begin{itemize}[leftmargin=*, topsep=1pt, itemsep=1pt]
    \item{\textbf{RQ1:} \textit{How effective are PLMs for fixing API misuses in methods?}}
    With the first RQ, we aim to examine the effectiveness of PLMs regarding their capability to repair API misuses in the method granularity.
    We are interested in investigating this RQ to ascertain the possibility of integrating PLM-based learning-aided APR techniques directly into real development.
    \item{\textbf{RQ2:} \textit{How do PLMs and specific APR techniques perform in fixing API misuses in methods with single-line changes?}}
    With the second RQ, we investigate the effectiveness of learning-aided APR techniques in repairing API misuses.
    We run two state-of-the-art APR techniques, i.e., SequenceR~\cite{sequencer}, Recoder~\cite{recoder}.
    Since SequenceR and Recoder are designed to repair single-line and single-hunk bugs, respectively, we further derive a smaller dataset (named \singledata) containing only single-line bugs, which can be fixed by modifying one line.
    Thus, we train/fine-tune and evaluate all the models on \singledata.
\end{itemize}

\subsection{Dataset}
\label{sec:benchmark}
\begin{figure}[t]
\centering
\renewcommand{\arraystretch}{1}

\begin{lstlisting}[language=json-pretty,style=Highlight]
{
   "Line":"=> 29",
   "parseTypeFail":"success",
   "Pattern":"UNKNOWN=>LocalVariable",
   "RetCheckAPI":[],
   "Type":"Insert",
   "BugDetectionTag":"[]",
   "Content":"=> int counter = 0",
   "FileName":"/home/PATH/BugDetectionProject/3Clone/NewDown2011-2017/All/V9/8530/buggy-version/jTowerDefense.src.model.Path.java",
   "BodyUseAPI":[],
   "Fixed commit":"96257b05bd27ed7a3d3ba55c80cf40de6aebb015",
   "Url":"https://api.github.com/repos/bernhardfritz/jTowerDefense",
   "Date":"2014-04-02T19:51:51Z"
}
\end{lstlisting}

\caption{An example of a data point in \textit{Li et al.'s data}.}
\label{fig:icst-data-example}
\end{figure}

In this section, we describe how we build the benchmark dataset.
We leverage the dataset published by Li et al.~\cite{li2021large} (\textit{Li et al.'s data}).
Figure~\ref{fig:icst-data-example} shows one data point in \textit{Li et al.'s data}.
We exclude the data points where (1) buggy line or fixed line is missing in \texttt{Line}, (2) \texttt{parseTypeFail} is not equal to \texttt{success}, or (3) the \texttt{Pattern} contains \texttt{UNKNOWN}. In this example, the buggy line is empty, and the \texttt{Pattern} contains \texttt{UNKNOWN}; thus, we exclude this data point from our dataset.
Although the dataset contains the commits that fixed API misuses (\texttt{Fixed commit}), it lacks the commits that contain the bugs.
We refer to these two types of commits as \textit{fixed} commits, and \textit{buggy} commits, respectively.
We extracted buggy commits by referring to the parent commits of the fixed commits.
Next, we downloaded the Java files of the two versions.
We used JavaParser~\footnote{\url{https://github.com/javaparser/javaparser}} to remove comments and parse the Java files.
We removed Java files that JavaParser cannot parse.
Similar to the prior work~\cite{tufano2019empirical}, we focus on the method granularity in this work.
Based on the \texttt{Line} information provided in Li et al.'s dataset, we extracted the buggy and fixed method pairs.
We further removed the duplicate method pairs, which may come from different forks while having the same base repository.
Finally, we shuffled the dataset and split the dataset into training, validation, and testing sets with a ratio of 8:1:1, as the \textit{complete data}.
To get the \singledata, we filter the bugs which only involve one-line change.
We also split them into training, validation, and testing with the same ratio.
For PLMs, the training and validation data are used for fine-tuning the models.
Table~\ref{tab:dataset} shows the statistics of our dataset, i.e., \textit{complete data} and \singledata.

\begin{table}[t]
    \centering
    \small
    \caption{Final dataset: the number of method pairs in each set and the average number of tokens in the src (buggy method) and tgt (fixed method).}
    \label{tab:dataset}
    \begin{tabular}{|cccccc|}
    \hline
    \multirow{2}{*}{\textbf{Variant}} & 
    \multirow{2}{*}{\textbf{Train.}} & 
    \multirow{2}{*}{\textbf{Valid.}} & 
    \multirow{2}{*}{\textbf{Test}} &
    \multicolumn{2}{c|}{\textbf{Avg. \# Tokens}} \\
    \cline{5-6}
    & & & & \textbf{SRC} & \textbf{TGT} \\
    \hline
    \textit{complete data} & 94,792 & 11,849 & 11,849 & 112 & 115 \\
    \hline
    \textit{single-line data} & 43,608 & 5,451 & 5,451 & 90 & 90 \\
    \hline
    \end{tabular}
\end{table}

\begin{table}[t]
    \centering
    \small
    \caption{Comparison of the selected \textit{general-purpose} PLMs.}
    \label{tab:ptms}
    \begin{tabular}{|lcc|}
    \hline
    \textbf{Model}\xspace & \textbf{Arch.} & \textbf{\# of Parameters} \\
    \hline
    CodeBERT~\cite{codebert} & Enc. & 125M \\
    
    GraphCodeBERT~\cite{graphcodebert} & Enc. & 125M \\
    \hline
    CodeGPT~\cite{codegpt} & Dec. & 124M \\
    
    PolyCoder-160M~\cite{polycoder} & Dec. & 160M \\
    
    PolyCoder-0.4B~\cite{polycoder} & Dec. &  400M \\
    \hline
    CodeTrans~\cite{codetrans} & Enc. \& Dec. & 223M \\

    PLBART~\cite{plbart} & Enc. \& Dec. & 140M \\
    
    CodeT5~\cite{codet5} & Enc. \& Dec. & 223M \\
    
    UniXCoder~\cite{unixcoder} & Enc. \& Dec. & 125M \\
    \hline
    \end{tabular}
\end{table}

\subsection{Selected Approaches}
Learning-aided APR approach takes a code snippet with API misuse(s) as the input sequence, and generates a fixed version of this code snippet that does not contain API misuse(s) as the output sequence.
We consider two types of learning-aided APR approaches, i.e., \textit{general-purpose} pre-trained models of code, which can be used to solve several programming understanding and generation tasks~\cite{zeng2022extensive,codegpt}, and the \textit{specialized techniques}, which are proposed to repair program errors.

\textbf{General-purpose Pre-trained Models of Code.}
PLMs mainly differ in architecture and pre-training tasks.
We include the PLMs that adopt three types of architecture, i.e., encoder, decoder, and encoder-decoder.
For each type of architecture, we choose state-of-the-art models that have demonstrated effectiveness in a recent study on programming understanding and programming generation tasks~\cite{zeng2022extensive}.
We also consider the more recent PLMs, i.e., PolyCoder~\cite{polycoder} and UniXCoder~\cite{unixcoder}.
Table~\ref{tab:ptms} shows the considered PLMs' architecture and the number of parameters.

\begin{itemize}[leftmargin=*, topsep=1pt, itemsep=1pt]
\item{\textbf{CodeBERT~\cite{codebert}}} is a bi-modal pre-trained model for programming language (PL) and natural language (NL).
It has been pre-trained with a hybrid objective function, including standard masked language modeling (MLM)~\cite{kenton2019bert} and replaced token dection~\cite{clark2019electra}.
Simply put, the MLM objective is to predict the original tokens which are masked out.
The replaced token detection objective predicts whether a token is an original token or not.
In the pre-training phase, the input is the concatenation of NL text and code in a certain PL with a special token to separate them.
CodeBERT considers both NL and PL as a sequence of words.
Specifically, the pre-training corpus of CodeBERT is a recent dataset CodeSearchNet ~\cite{husain2019codesearchnet}, which contains 2.1M bimodal data points and 6.4M unimodal data points.

\item{\textbf{GraphCodeBERT~\cite{graphcodebert}}} considers the inherent structure of code.
In the pre-training stage, it uses data flow, a semantic-level structure of code that encodes the relationship between variables.
Other than MLM, GraphCodeBERT also adopts two structure-aware pre-training tasks: one is data flow edge prediction, which aims to learn representation from code structure; the other is to align representations between source code and code structure.
GraphCodeBERT was pre-trained on the CodeSearchNet dataset~\cite{husain2019codesearchnet}, which contains
2.4M functions of six programming languages paired with natural language documents. 

\item{\textbf{CodeGPT~\cite{codegpt}}} has the same model architecture and pre-training objective as GPT-2~\cite{radford2019language}.
Specifically, CodeGPT has 12 layers of Transformer decoders.
GPT-2 is trained with a simple objective: predicting the next token one by one, which is conditioned on its previous tokens and itself.
This is also called auto-regressive language modeling.
The one we evaluated in our work is the \textit{CodeGPT-adapted} variant, which uses GPT-2 as the starting point and is further pre-trained in the 1.6M Java methods from CodeSearchNet dataset~\cite{husain2019codesearchnet}.
It has the same vocabulary and natural language understanding ability as the original GPT-2.

\item{\textbf{PolyCoder~\cite{polycoder}}} is also based on GPT-2 structure~\cite{radford2019language} and was pre-trained on the dataset collected by its authors.
After deduplication and filtering, its pre-training dataset contains 24.1M files and 254GB of data across 12 PLs.
In the original work, the authors trained three models in different sizes, with 2.7 billion, 400 million, and 160 million parameters.
Considering our budget, in our work, we choose the last two models, which have 160M and 400M parameters, respectively.
We leave the evaluation of a larger variant, i.e., the 2.7B-parameter model, for future work.

\item{\textbf{CodeTrans~\cite{codetrans}}} is based on the T5 architecture~\cite{raffel2020exploring}.
In the pre-training stage, CodeTrans involved six different corpora for unlabeled datasets, which cover 9 PLs and English text.
In total, it has around 40 million samples.
It applies the span corruption task with a corruption rate of 15\% as the pre-training task.
It corrupts the input sequence by masking a span of tokens, and then the model is trained to predict the masked spans.

\item{\textbf{PLBART~\cite{plbart}}} is a bidirectional and autoregressive Trans-
former pre-trained on unlabeled data across PL and NL.
PLBART uses the same architecture as BART$_{base}$~\cite{lewis2020bart}: it has 6 layers of an encoder and 6 layers of a decoder.
Similar to CodeTrans, PLBART also uses denoising Seq2Seq pre-training: the model learns to reconstruct an input text that is corrupted by a noise function.
Specifically, PLBART adopts three noising strategies: token masking, deletion, and infilling.
PLBART has been pre-trained on a large collection of Java and Python functions
and their NL descriptions from GitHub and Stack Overflow. 

\item{\textbf{CodeT5~\cite{codet5}}} also builds upon the T5 architecture~\cite{raffel2020exploring}, while it considers the token type information in code.
Similarly, CodeT5 employs a denoising sequence-to-sequence pre-training task.
Moreover, CodeT5 leverages the code semantics conveyed by the developer-assigned identifiers.
The model is also pre-trained with two identifier-related tasks.
The first task is \textit{identifier tagging}, and aims to distinguish whether the code token is an identifier or not.
The other task is \textit{masked identifier prediction}, which corrupts the input sequence by masking all the identifiers in the PL segment and employs a sentinel token for all occurrences of one specific identifier.

\item{\textbf{UniXCoder~\cite{unixcoder}}} is a unified cross-modal pre-trained
model for PL. 
Its pre-training tasks are MLM, unidirectional language modeling, and the denoising objective.
After these three pre-training tasks, to learn the semantic embedding, UniXCoder proposes two pre-training tasks, i.e., multi-modal contrastive learning and cross-modal generation.
For multi-modal contrastive learning, positive examples are the same input with different hidden dropout masks, and negative examples are other representations in the same batch.
For the cross-modal generation, the model needs to generate a comment describing the function of the code.
UniXCoder supports three modes: \textit{encoder}, \textit{decoder}, and \textit{encoder-decoder}.
We use the encoder-decoder mode of UniXCoder.

\end{itemize}

\textbf{Specialized Techniques.}
\label{sec:spec-tools}
To select specialized APR techniques, we mainly rely on the latest live review about automatic program repair~\cite{monperrus2018living}, which was submitted on 9 Aug 2022.
We mainly focus on the general data-driven APR approaches proposed in the last three years (2020 - 2022).
Thus, we do not consider the domain-special APR techniques, such as APR for null pointer error~\cite{lee2022npex} and APR for concurrency errors~\cite{li2019dfix}.
The criteria to be selected for inclusion in our study are (1) a new APR approach has been proposed, (2) the source code is publicly available, and (3) the authors provide an easy-to-follow guide to reproduce their work so that we can adapt the proposed approach to a new dataset.
Note that it is not a trivial task to replicate these APR techniques, as they are often implemented in different deep-learning libraries and require different dependencies.
Existing APR techniques usually compare with each other in Defects4J~\cite{just2014defects4j}; thus, they often directly cite the results of prior reported results instead of re-running the experiments~\cite{xia2022less,cure,liu2019tbar,coconut,recoder}.
Furthermore, given the large size of our dataset, it is infeasible to run test suites for each potential fix constantly.
Thus, we exclude APR techniques that rely on test suites (either in the training or inference stage), e.g., RewardRepair~\cite{ye2022neural} and SelfAPR~\cite{selfapr} involve test cases in the training stage, and DEAR~\cite{li2022dear} requires test cases in the inference stage.
Given we already have a group of PLMs, we excluded the APR techniques which rely on PLMs, such as CURE~\cite{cure}.
In the end, we choose one Seq2Seq model, i.e., SequenceR~\cite{sequencer}, and one tree-based model, i.e., Recoder~\cite{recoder}.

\begin{itemize}[leftmargin=*, topsep=1pt, itemsep=1pt]
    \item{\textbf{SequenceR~\cite{sequencer}}} is based on Seq2Seq learning and it adopts copy mechanism~\cite{see2017get} to overcome the unlimited vocabulary issue in source code. 
    It is specifically designed to solve one-line patch generation task, i.e., the bug can be fixed by replacing a single line.
    \item{\textbf{Recoder~\cite{recoder}}} is built based on encoder-decoder architecture. It uses a syntax-guided edit decoder with placeholder generation, aiming to generate a sequence of edits rather than a new statement. The usage of this decoder tackles the problem of inefficient representation of small edits.
    Moreover, it also enables Recoder to generate project-specific identifiers by leveraging a neural network for placeholder generation.
    Recoder leverages AST in the framework. It first transforms the raw method into an AST, which is then modified and used to extract the rules. The generated rule will be used in the training and inference phase.
\end{itemize}

\subsection{Implementation}
\textbf{PLM.} We implement PLMs with Hugging Face \texttt{Transformers} library~\cite{wolf-etal-2020-transformers}.
We run each model for 30 epochs under each setting.
However, if the loss in the validation set does not decrease for 5 epochs, the early-stopping strategy would be triggered.
We used the model which has the smallest loss on the validation set as the final model.
The used hyper-parameters are available in our replication package~\footnote{\url{https://anonymous.4open.science/r/TOSEM-API-Misuse}}.
We run the experiments on a machine with 4 NVIDIA RTX A5000 GPUs and the AMD EPYC 7643 48-Core Processor.

\textbf{APR techniques.}
We set a beam size of five for both approaches.
For SequenceR, we keep the Top 1 prediction result as the PLMs.
We used the model which achieves the highest accuracy on the validation set as the final model.
For Recoder, we set the batch size to 16, and the epoch to 30 for the training phase.
In the inferring phase, we keep the Top 1 prediction result as the fix for the buggy code.
As for other hyper-parameters, we used the default values as provided in the replication package.

To fairly compare the performance of PLMs and the studied APR techniques, we give the same information to all the models, i.e., the input is the buggy method.
Even so, the specific input format may vary across different models.
Based on the original design, for SequenceR~\cite{sequencer}, we provide \texttt{<START\_BUG>} and \texttt{<END\_BUG>} labels to explicitly separate the buggy lines and the context in the buggy method.
As for Recoder~\cite{recoder}, we preprocess the raw buggy method with a script provided in Recoder's replication package to prepare the input data.
In general, it takes the raw buggy method as input in the training phase. 
For the inference phase, we need to provide the location of the buggy line along with the buggy method.
On the other hand, the rest models were given the raw buggy method as the input.

\subsection{Evaluation Metrics}
Following prior work~\cite{codet5,lee2022npex}, we use the Exact Match (EM), BLEU~\cite{papineni2002bleu}, and CodeBLEU~\cite{ren2020codebleu} to measure the quality of the generated fixed code.
We consider EM as the critical metric, as it is the strictest one.
We also consider BLEU, since it is the widely-used metric for NMT.
On the other hand, CodeBLEU considers the unique syntax and semantics of source code.
Moreover, compared with BLEU and EM, it shows a better correlation with the programmer-assigned scores~\cite{ren2020codebleu}.
We describe each metric as follows.

\textit{EM (Exact Match)} is defined as the percentage of the generated fixed code that \textit{exactly matches} the reference fixed code.
It is a strict metric that can exclude some successful repairs.
EM can be considered as the lower bound since the different programs can be semantically the same but are written differently from the developer-written code.

\textit{BLEU-4}, calculates the percentage of 4-gram overlap between the \textit{reference} fix code and the \textit{candidate} code.
For simplicity, we refer it as \textit{BLEU} in our work.
Candidate fixed code represents the code generated by models, while reference fixed code represents the ground-truth code written by developers.
BLEU is defined as Equation~\ref{eq:bleu-4}:

\begin{equation}
\label{eq:bleu-4}
\text{BLEU}=\frac{\sum_{P \in C} \sum_{4-gram \in P} \text{Count}_{\text{matched}}(\text{4-gram})}{\sum_{P \in C} \sum_{4-gram \in P} \text{Count}(\text{4-gram})}
\end{equation}

\noindent, where $P$ refers to each candidate, fix code generated by the model, $C$ refers to the whole candidate code.
Since it was proposed for evaluating NL, it neglects the syntax and semantics included in the source code.

To remedy the drawback of applying BLEU in source code tokens, we include the newly introduced metric \textit{CodeBLEU}.
CodeBLEU considers both syntactic match and semantic match by injecting code syntax via AST and code semantics via data flow.
CodeBLEU is defined in Equation~\ref{eq:codebelu}:

\begin{equation}
\label{eq:codebelu}
\begin{aligned}
&\operatorname{CodeBLEU}=\alpha \times \mathrm{BLEU}+\beta \times \mathrm{BLEU}_{\text {weight }}\\
&+\gamma \times \operatorname{Match}_{\text {ast }}+\delta \times \operatorname{Match}_{d f}
\end{aligned}
\end{equation}

\noindent where $BLEU_{weight}$ is the weighted $n$-gram match, $Match_{ast}$ is the syntactic AST match, $Match_{df}$ is the semantic data-flow match.
Both $BLEU$ and $BLEU_{weight}$ work in sequence-level matching, while the latter one considers the keywords in each PL.
$Match_{ast}$ calculates the accuracy by comparing the sub-trees from both candidate and reference code.
$Match_{df}$ computes the semantic data-flow match score between the candidate and reference code.
In our work, we use the default values of $\alpha$, $\beta$, $\gamma$, and $\delta$ as 0.25.
\section{Results}
\label{sec:result}
\subsection{RQ1: How effective are PLMs for fixing API misuses in methods?}
\begin{table}[t]
    \centering
    \small
    \caption{Results of PLMs in repairing API misuse on \textit{complete data}. The best performer in each group is in bold.}
    \label{tab:rq1}
    \begin{tabular}{|lccc|}
    \hline
   \textbf{Model} & \textbf{BLEU} & \textbf{CodeBLEU} & \textbf{EM}  \\
    \hline
    \underline{\textit{Encoder-based}} & & & \\
    CodeBERT~\cite{codebert} &  \textbf{51.7}	& \textbf{58.22}	& 1 \\
    
    GraphCodeBERT~\cite{graphcodebert} & 49.21	& 56.25	& \textbf{1.55}  \\
    \hline
    \underline{\textit{Decoder-based}}  & & & \\
    
    CodeGPT~\cite{codegpt} & 50.85 & 	57.57& 	2.02 \\
    
    PolyCoder-160M~\cite{polycoder} & 54.58	& 60.36	& 2.77 \\
    
    PolyCoder-0.4B~\cite{polycoder} & \textbf{54.88} & 	\textbf{60.62} & 	\textbf{2.96} \\
    \hline
    \underline{\textit{Encoder-decoder-based}}  & & & \\
    CodeTrans~\cite{codetrans} & 53.12	& 59.24	& 2.68 \\
    
    PLBART~\cite{plbart} & \textbf{54.9}	& 60.15	& 1.34 \\
    CodeT5~\cite{codet5} & 54.08	& 60 & 	\textbf{3.01} \\
    
    UniXCoder~\cite{unixcoder} & 54.76	& \textbf{60.36}	& 2.17 \\
    \hline
    \end{tabular}
\end{table}

\textbf{Quantitative Analysis.}
Table~\ref{tab:rq1} shows the performance of the three groups of fine-tuned PLMs.
We can find that decoder-based and encoder-decoder-based PLMs generally perform better than encoder-based PLMs, except that PLBART achieves a lower EM ratio than GraphCodeBERT.
Based on the task characteristics, the decoder-based and the encoder-decoder-based PLMs are more suitable for APR.
Specifically, the pre-training tasks adopted by the encoder-decoder group resemble our API misuse repair task the most.
It suggests that the Seq2Seq pre-training can benefit the downstream Seq2Seq task, where in our case, the downstream task is repairing API misuse.

The best-performing encoder-decoder-based model, CodeT5, performs similarly to the best-performing decoder-based model, i.e., PolyCoder-0.4B.
Still, CodeT5 outperforms PolyCoder-0.4B by 1.7\% in terms of EM.
Besides, PolyCoder-0.4B and CodeT5 achieve similar BLEU and CodeBLEU scores.
Moreover, all the four encoder-decoder-based models achieve similar BLEU and CodeBLEU scores.
This suggests that the repairs generated by encoder-decoder-based PLMs are likely to be syntactically correct or semantically similar to the developer-written repair.
The gap between CodeT5 and the best-performing encoder-based model, i.e., GraphCodeBERT, is more pronounced.
According to our main metric, i.e, EM, CodeT5 outperforms GraphCodeBERT by 94.2\%.
For the other two metrics, CodeT5 also achieves higher values.

We also compare the performance of PLMs within the same group.
The result shows that PolyCoder can achieve a higher EM than CodeGPT.
It suggests that the larger the pre-training corpora, the model tends to perform better.
PolyCoder and CodeGPT are based on the GPT-2 architecture, while PolyCoder has been pre-trained in a larger corpus than CodeGPT. 
Similarly, the better performance of PolyCoder-0.4B over PolyCoder-160M indicates that, with the same model architecture and the same pre-training task, the one having more parameters tends to achieve better performance.

\begin{figure}[t]
    \centering
    \includegraphics[width=0.5\textwidth]{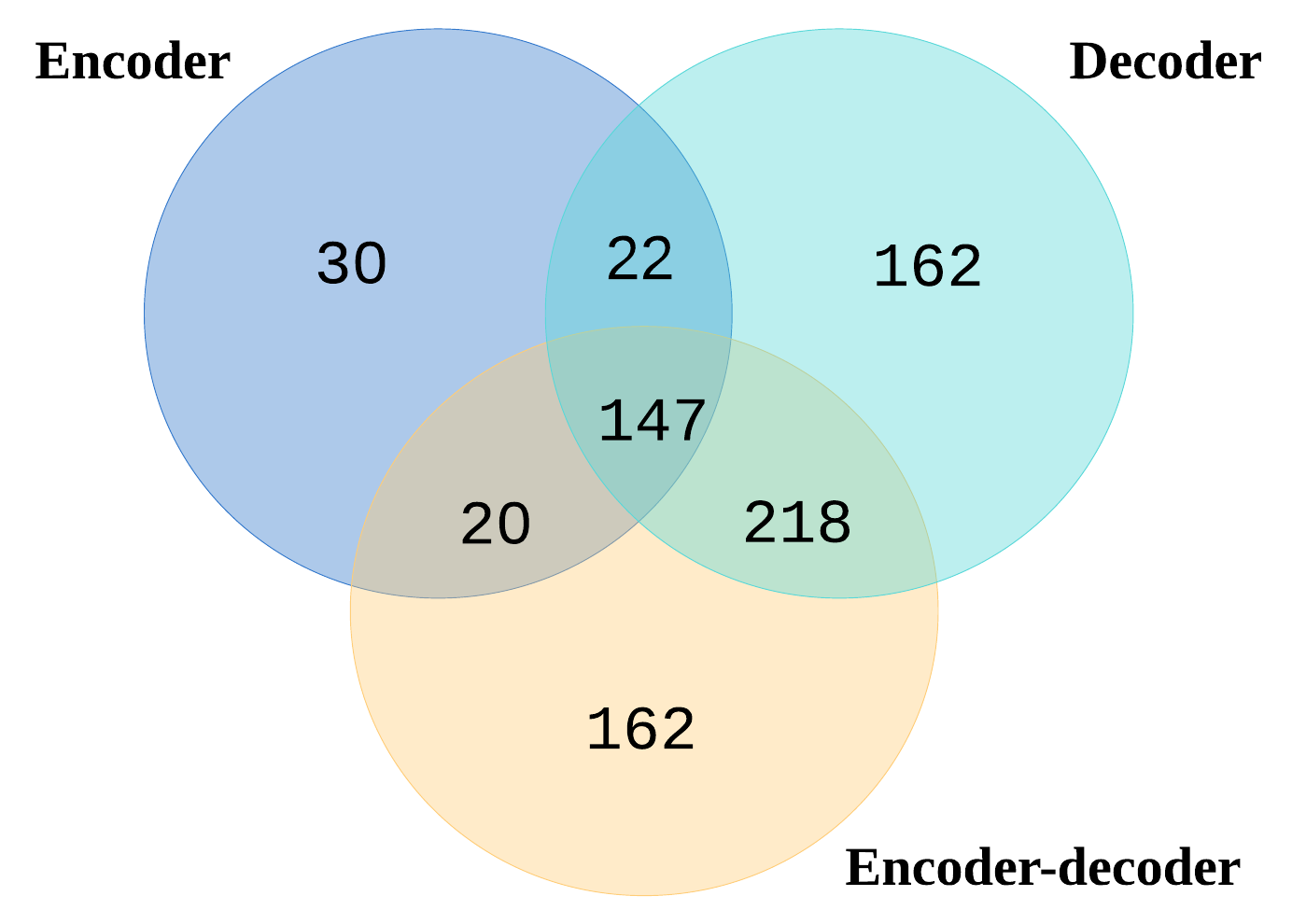}
    \caption{The number of successful repairs produced by different groups of PLMs in the \textit{complete data}.}
    \label{fig:complete}
\end{figure}

We merge the successful repair produced by the PLMs inside the same group. 
We consider the repair to be \textit{successful} if the generated repair is exactly the same as the developer-written repair.
Figure~\ref{fig:complete} shows the overlap among the successful repairs produced by different groups of PLMs.
We can find that fine-tuned decoder-based and encoder-decoder-based models can produce more successful repairs than encoder-based models.
We further analyze the methods that each PLM group has successfully repaired.
All three groups of PLMs successfully repair the same 147 buggy methods.
We find that each group of PLM can successfully generate the correct repairs for some buggy methods that the other two groups cannot.
We also observe that any two groups of PLMs have common methods that they both successfully repair, but the remaining group cannot.
In total, all the PLMs can repair 761 buggy methods, which accounts for 6.42\% of all the method pairs.

\textbf{Qualitative Analysis.}
To understand the reason behind the performance difference between these groups of PLMs, we analyze the generated repairs.
Specifically, we performed a qualitative analysis to understand when the APR techniques can fix the API misuse, and when they cannot.
To understand why they can fix API misuses, we investigate the common successful repairs produced by the three groups of PLMs.
We manually check the successful repairs by all three groups of PLMs (i.e., 147 repairs).
Two authors independently categorized all 147 buggy-fixed method pairs into three mutually exclusive categories. 
These categories were based on whether the fixes could be generated by: (1) Understanding method semantics, (2) Changing standard Java APIs, or (3) Others, which encompassed fixes requiring information beyond the buggy method. 
The authors then discussed and resolved any disagreements. 
The resulting categories consisted of 77 pairs (52.4\%), 13 pairs (8.8\%), and 57 pairs (38.8\%), respectively.
We show two examples in Figure~\ref{fig:rq1-scs}, where lines starting with `-' are the buggy lines, and those starting with `+' are the correct fixes written by developers. 
We have the following findings.

\begin{figure}[t]
	\centering
	\renewcommand{\lstlistingname}{Method}
	\lstset{
 language=java,
  numbers=left,
  frame=shadowbox,
  rulesepcolor= \color{red!20!green!20!blue!20},
  basicstyle=\footnotesize\ttfamily,
  numberstyle=\scriptsize,
  breaklines=true,
  columns=fullflexible,
  xleftmargin=16pt,
  showstringspaces=false,
  keywordstyle=\color{blue}\bfseries,
		commentstyle= \color{red!50!green!50!blue!50},
        xrightmargin=0em, aboveskip=0.5em,
		framexleftmargin=1.5em,
        numbersep= 5pt,
        emphstyle=\bfseries,
        moredelim=**[is][\color{red}]{@}{@},
		escapeinside= {(*@}{@*)}
}
\renewcommand{\arraystretch}{1}
\begin{flushleft}
\underline{\small \textbf{Example 1:} Inferred based on the semantics of the identifier} \\
\end{flushleft}

\begin{lstlisting}
 private Period clonePeriodAfterMidnight(Period source) { 
    Period result = new Period(); 
    result.setStartDate(source.getStartDate().plusDays(1));
(*@{\color{red}{-\quad result.setEndDate(source.\textbf{getStartDate()}.plusDays(1)); }@*)
(*@{\color{javagreen}{+\quad result.setEndDate(source.\textbf{getEndDate()}.plusDays(1)); }@*)
    return result; 
}
\end{lstlisting}

\begin{flushleft}
\underline{\small \textbf{Example 2:} Standard Java API} \\
\end{flushleft}
\begin{lstlisting}
public boolean containsKey(Object key) {
(*@{\color{red}{-\quad  return this.parametersHashTable.\textbf{contains}(key);  }@*)
(*@{\color{javagreen}{+\quad return this.parametersHashTable.\textbf{containsKey}(key); }@*)
}
\end{lstlisting}

\caption{Common successful fixes generated by PLMs}
\label{fig:rq1-scs}
\end{figure}

\begin{enumerate}[leftmargin=*]
    \item \textit{PLMs can understand the method semantics.}
    We identify several fixes related to correcting the use of API calls.
    When the API name conveys meaningful information or some existing statements as the context, PLMs can identify statements that are inconsistent with the method context.
    Consider Example 1 in Figure~\ref{fig:rq1-scs},
    inside the definition of method \texttt{clonePeriodAfterMidnight}, there are four statements.    
    Line 3 sets the start date. Naturally, Line 4 should set the end date.
    The API call was correct, i.e., \texttt{result.setEndDate()}, but the argument, which is filled with another API call was wrong: \texttt{getStartDate()} is called instead of \texttt{getEndDate()}. 
    Based on the method name and the context of the method, PLMs can learn to fix this type of bug.
    
    \item \textit{PLMs can repair the misuse of standard Java APIs.}
    We find that PLMs can repair the APIs in the standard Java libraries.
    For Example 2 in Figure~\ref{fig:rq1-scs}, both \texttt{contains} and \texttt{containsKey} are correct APIs in \texttt{HashTable}.
    However, the arguments are different.
    \texttt{contains} checks whether some key maps into the specified \textit{value} in this HashTable; while \texttt{containsKey} checks whether a specified object is a \textit{key} in this HashTable.
    Based on the parameter name, the function aims to check whether the key exists.
\end{enumerate}

To understand why PLMs fail to generate the correct repair, we randomly sampled 100 methods that all PLMs fail to repair.
The aim is to determine the challenges of repairing API misuse with PLMs.
We found several cases where the current PLMs fail.

\begin{figure}[t]
	\centering
	\lstset{
 language=java,
  numbers=left,
  frame=shadowbox,
  rulesepcolor= \color{red!20!green!20!blue!20},
  basicstyle=\footnotesize\ttfamily,
  numberstyle=\scriptsize,
  breaklines=true,
  columns=fullflexible,
  xleftmargin=16pt,
  showstringspaces=false,
  keywordstyle=\color{blue}\bfseries,
		commentstyle= \color{red!50!green!50!blue!50},
        xrightmargin=0em, aboveskip=0.5em,
		framexleftmargin=1.5em,
        numbersep= 5pt,
        emphstyle=\bfseries,
        moredelim=**[is][\color{red}]{@}{@},
		escapeinside= {(*@}{@*)}
}
\begin{flushleft}
\underline{\small \textbf{Example 3:} Partial fix (Developer-written Fix)} \\
\end{flushleft}

\begin{lstlisting}
(*@ @ @*)Override 
public boolean isValid(Object value) { 
(*@{\color{red}{-\quad    \textbf{if (value != null) \{} }@*)
(*@{\color{red}{-\quad\quad\quad      \textbf{ value = "";}  }@*)
(*@{\color{red}{-\quad   \textbf{\}} }@*)
(*@{\color{javagreen}{+\quad \textbf{value = value.toString();} }@*)
    return mPattern.matcher((String) value).matches(); 
}
\end{lstlisting}

\begin{flushleft}
\underline{\small Repair generated by CodeT5} \\
\end{flushleft}

\begin{lstlisting}
(*@ @ @*)Override 
public boolean isValid(Object value) { 
(*@{\color{red}{-\quad  if (value \textbf{!=} null) \{ }@*)
(*@{\color{red}{-\quad\quad\quad    \textbf{value = "";}  }@*)
(*@{\color{javagreen}{+\quad if (value \textbf{==} null) \{ }@*)
(*@{\color{javagreen}{+\quad\quad\quad \textbf{return false;} }@*)
    }
    return mPattern.matcher((String) value).matches(); 
}
\end{lstlisting}

\begin{flushleft}
\underline{\small \textbf{Example 4:} Lack of context} \\
\end{flushleft}
\begin{lstlisting}
private void updatePTTConfiguration() {
(*@{\color{red}{-\quad \textbf{mTalkButton}.setVisibility(settings.isPushToTalk() \&\& settings.isPushToTalkButtonShown() ? View.VISIBLE : View.GONE); }@*)
(*@{\color{javagreen}{+\quad \textbf{pttView}.setVisibility(settings.isPushToTalk() \&\& settings.isPushToTalkButtonShown() ? View.VISIBLE : View.GONE); }@*)
}
\end{lstlisting}

\begin{flushleft}
\underline{\small Repair generated by CodeBERT} \\
\end{flushleft}

\begin{lstlisting}
private void updatePTTConfiguration() {
(*@{\color{red}{-\quad mTalkButton.setVisibility(\textbf{settings.isPushToTalk() \&\& settings.isPushToTalkButtonShown() ? View.VISIBLE : View.GONE}); }@*)
(*@{\color{javagreen}{+\quad mTalkButton.setVisibility(\textbf{true}); }@*)
}
\end{lstlisting}

\caption{Wrong fixes generated by PLMs}
\label{fig:rq1-failure}
\end{figure}

\begin{enumerate}[leftmargin=*]
\item \textit{When failing to consider the impact of its changes.}
In some cases, PLMs provide a partial fix.
Take Example 3 in Figure~\ref{fig:rq1-failure}: in the buggy method, if the parameter \texttt{value} is not null, the \texttt{value} would be set as an empty string.
As a result, the argument of the API call in Line 7 would always be the same.
It is certainly not the intention of this method.
The fix generated by CodeT5 makes more sense: if the \texttt{value} is \texttt{null}, return \texttt{false}; otherwise, make the API call in Line 7.
However, it neglects the fact that the String casting only works if the Object is an instance of String.
Since \texttt{value} can be any Object, using \texttt{(String)} is erroneous. 
The fix provided by a developer handles this issue.

\item \textit{When the intention is unclear.}
In some cases, without knowing the developer's intention, both the buggy and fixed versions may look bug-free.
As shown by Example 4 in Figure~\ref{fig:rq1-failure}, the buggy method and the fixed method differ in the internal class property being called, i.e., either a button or a view.
Without knowing the exact intention and what \texttt{ptt} refers to, it is nearly impossible to fix this kind of bug.
Given the \textit{buggy} version can be considered correct, the generated repair by CodeT5 and PolyCoder are the same as the buggy version. 
On the other hand, CodeBERT replaces the parameter with \texttt{true}.
\end{enumerate}

\begin{tcolorbox}[left=4pt,right=4pt,top=2pt,bottom=2pt,boxrule=0.5pt]
\textbf{Answer to RQ1:} 
PLMs are effective in fixing API misuses in methods.
Specifically, decoder-based and encoder-decoder-based PLMs perform better than encoder-based PLMs.
\end{tcolorbox}

\subsection{RQ2: How do PLMs and specific APR techniques perform in fixing API misuses in methods with single-line changes?}

\begin{table}[t]
    \centering
    \small
    \caption{Results of PLMs and APR techniques on the \textit{single-line data}.}
    \label{tab:rq2}
    \begin{tabular}{|lccc|}
    \hline
    \textbf{Model} & \textbf{BLEU} & \textbf{CodeBLEU} & \textbf{EM}  \\
    \hline        
    \underline{\textit{Encoder-based}} & & & \\
    CodeBERT~\cite{codebert} & \textbf{63.11} &	\textbf{66.58} & 2.18 \\
    
    GraphCodeBERT~\cite{graphcodebert} & 60.5 &	64.17	& \textbf{2.49} \\
    \hline
    \underline{\textit{Decoder-based}} & & & \\
    CodeGPT~\cite{codegpt} & 62.21 &	65.9	&4.22 \\
    
    PolyCoder-160M~\cite{polycoder} & 65.7&	\textbf{68.66}	&5.93 \\
    
    PolyCoder-0.4B~\cite{polycoder} & \textbf{65.73} &	68.65	& \textbf{6.31} \\
    \hline
    \underline{\textit{Encoder-decoder-based}}  & & &\\
    CodeTrans~\cite{codetrans} & 64.43	&67.47	&5.65 \\
    
    PLBART~\cite{plbart} &  \textbf{66.13} &	\textbf{68.8} &	5.1 \\

    CodeT5~\cite{codet5} & 65.18&	68.27	& \textbf{8.86} \\
    
    UniXCoder~\cite{unixcoder} & 66.01	&68.73	&4.97 \\

    \hline
    \hline
    SequenceR~\cite{sequencer} & 93.58 & 89.75 & 0.48 \\
    Recoder~\cite{recoder} & \textbf{94.05} & \textbf{93.4} & \textbf{0.72} \\
    \hline
    \end{tabular}
\end{table}

\textbf{Quantitative Analysis.}
Table~\ref{tab:rq2} shows the comparison of PLMs against the state-of-the-art APR techniques on the \singledata.

\begin{figure}[t]
    \centering
    \includegraphics[width=0.4\textwidth]{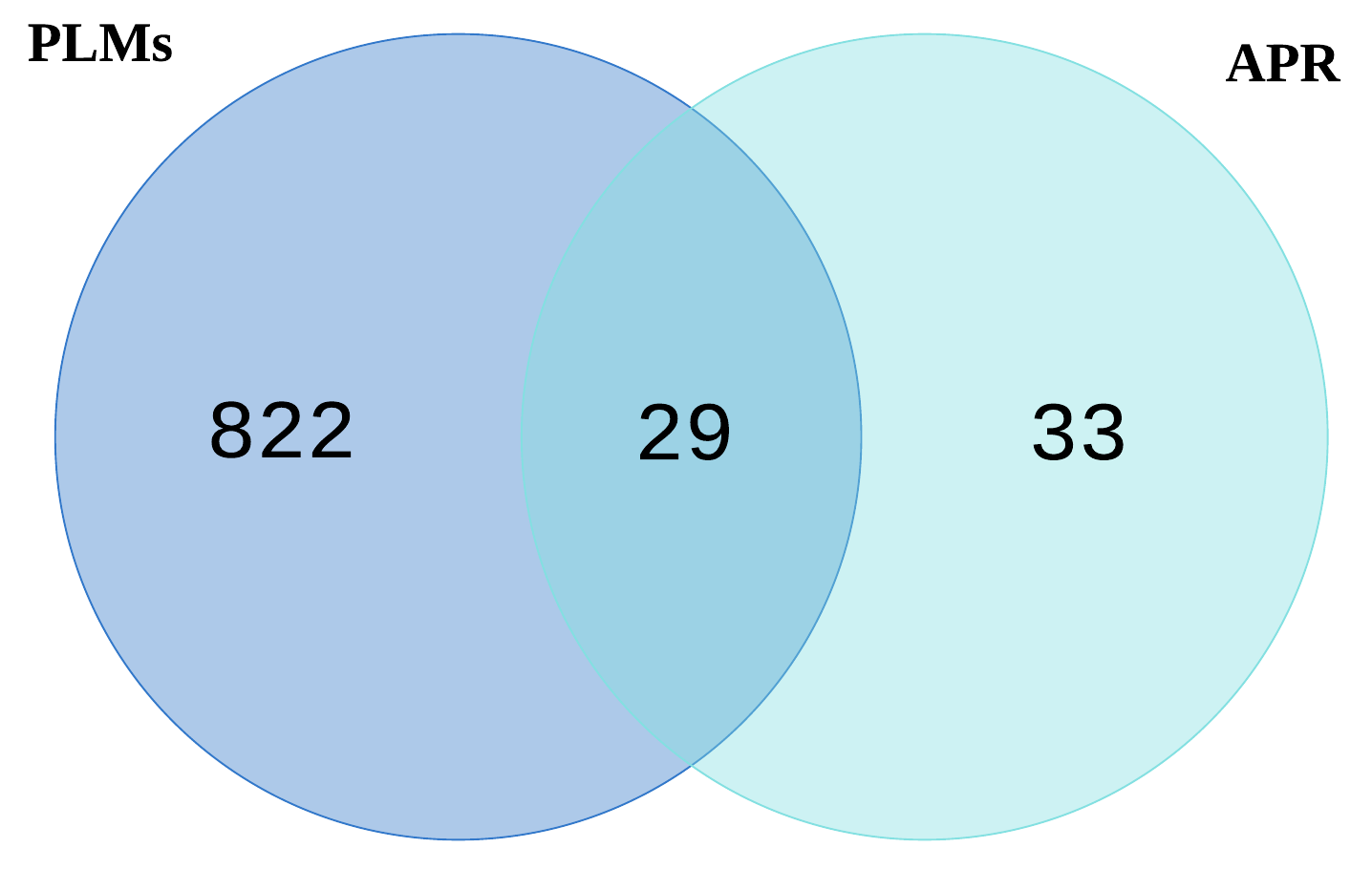}
    \caption{The number of successful repairs produced by PLMs and APR techniques in the \textit{single-line data}.}
    \label{fig:complete}
\end{figure}

Table~\ref{tab:rq2} shows that fine-tuned PLMs perform better than studied APR techniques, which demonstrates the effectiveness of pre-training: all of the PLMs are pre-trained in a large corpus to capture the syntactic and semantic knowledge of source code.
The best-performing model, CodeT5, outperforms Recoder by around 12 times in terms of EM.
SequenceR and Recoder perform worse than PLMs, with the worst-performing CodeBERT achieving more than double the EM ratio.
Regarding BLEU and CodeBLEU, it may be surprising to see that the studied APR approaches surpass the PLMs.
Simply copying the buggy method as the fixed method would result in a BLEU and CodeBLEU close to 100.
It is worth reminding that both selected APR approaches generate the fixed line instead of the whole method.
Thus, the large portion of the fixed method would be the same as the buggy method.
Therefore, it is intuitive that the APR approaches can achieve higher BLEU and CodeBLEU.

\textbf{Qualitative Analysis.}
In this section, we performed two sets of manual checks: 

1. Evaluation of the common successful fixes generated by PLMs and APR techniques. We thoroughly examined all 29 identified cases, which collectively represent the total number of fixes commonly identified across the various approaches.

2. Analysis of the characteristics of successful fixes produced solely by APR techniques. We carefully studied 33 cases where the APR tools successfully fixed the bugs, while the PLMs were unable to do so.

\begin{enumerate}[leftmargin=*]
\item \textit{Common successful fixes.} The fixes relate to the method argument. For example, the fix is to change the boolean argument in a method call: the fix is either changing \texttt{true} to \texttt{false} or vice versa. 
We found 11 out of 29 cases (37.9\%) were resolved in this manner.

\item \textit{Characteristic of unique successful fixes by APR techniques.} Similar to the prior finding, we found 12 out of 33 cases (36.4\%) were fixed by changing the boolean argument in the method call. The difference is that most cases involve larger methods with more lines than cases in the common successful fixes. This applies not only to the boolean argument fixes. We also find other successful fixes are generally longer, which makes it challenging to localize the API misuse and therefore results in PLMs failing to produce the correct fix.
Since PLMs lack explicit information on fault localization, APR techniques may be more aware of the faulty location. This suggests that an accurate bug localization may have benefited the APR techniques. 

\end{enumerate}

\begin{tcolorbox}[left=4pt,right=4pt,top=2pt,bottom=2pt,boxrule=0.5pt]
    \textbf{Answer to RQ2:} PLMs are more effective than the studied APR techniques in fixing API misuses in methods with single-line changes.
\end{tcolorbox}
\section{Discussion}
\label{sec:discussion}

\subsection{Impact of the Program Length on the Model Performance}
\label{sec:failure}
In Section~\ref{sec:result}, when we investigate the reasons why APR techniques fail to successfully produce the repairs, we noticed that the failure cases have relatively longer method lengths compared to the successful cases.
Therefore, we investigate whether the length of the buggy methods affects the model performance.

\begin{figure}[t]
    \centering
    \includegraphics[width=0.5\textwidth]{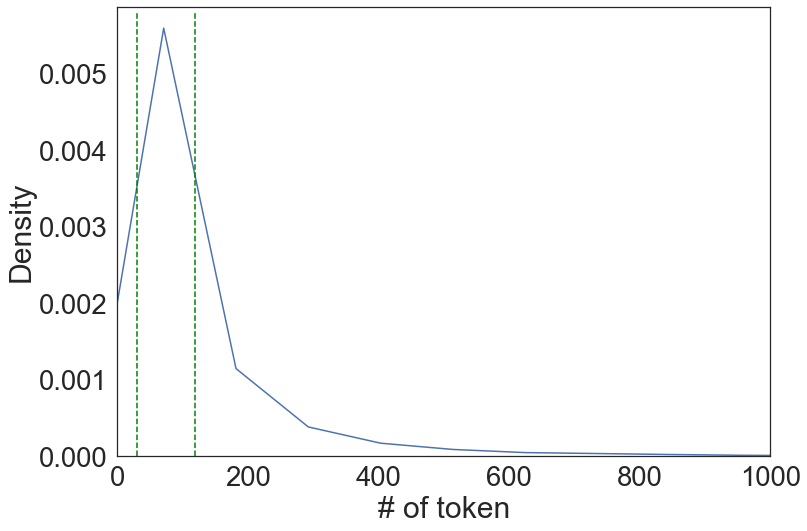}
    \caption{Distribution of buggy methods by the number of tokens in \textit{complete data}.}
    \label{fig:token-length}
\end{figure}

From Figure~\ref{fig:token-length}, we notice that there is a peak before 100 tokens and a long tail that extends over 300 tokens.
The first quartile for this distribution is 30 tokens, and the third quartile is 119 tokens.
We consider \textit{short} methods as buggy methods that have less or equal to 30 tokens and \textit{long} methods as buggy methods that have more than 119 tokens.
Since we want to focus on comparing the performance when the length difference of the methods is more pronounced, we only report the results with short and long methods.

\begin{table}[t]
    \centering
    \scriptsize
    \small
    \caption{Results on the \textit{short} (less or equal to 30 tokens) and \textit{long} (more than 119 tokens) method pairs in \textit{complete data}.}
    \label{tab:length-complete}
    \begin{tabular}{|lcccccc|}
    \hline
    & \multicolumn{2}{c}{\textbf{BLEU}} & \multicolumn{2}{c}{\textbf{CodeBLEU}}  & \multicolumn{2}{c|}{\textbf{EM}} \\
    \cline{2-7}
    & \textbf{short} & \textbf{long} & \textbf{short} & \textbf{long} & \textbf{short} & \textbf{long} \\
    \hline
    CodeBERT & \textbf{60.64} & \textbf{33.2} & \textbf{65.85} & \textbf{45.98} & 2.4 & 0.03 \\
    
    GraphCodeBERT & 60.22 & 30.3 & 65.56 & 43.52 & \textbf{3.5} & \textbf{0.17} \\
    \hline
    CodeGPT & \textbf{60.03} & 32.21 & \textbf{65.37} & 45.28 & 4.35 & 0.1 \\
    
    PolyCoder-160M & 58.29 & 38.42 & 63.89 & 49.68 & 5.87 & 0.27 \\
    
    PolyCoder-0.4B & 59.51 & \textbf{38.52} & 65.06 & \textbf{49.75} & \textbf{6.28} & \textbf{0.34} \\
    \hline
    CodeTrans & 61.12 & 35.32 & 66.33 & 47.45 & \textbf{5.46} & 0.34 \\
    
    PLBART & 60.91 & 38.08 & 65.95 & 49.01 & 2.71 & 0.21 \\
    
    CodeT5 & \textbf{61.17} & 36.74 & \textbf{66.39} & 48.47 & 6.15 & \textbf{0.38} \\
    
    UniXCoder & 60.14 & \textbf{38.62} & 65.65 & \textbf{49.61} & 4.95 & 0.03 \\
    \hline
    \end{tabular}
\end{table}

Table~\ref{tab:length-complete} shows the results of different PLMs on both \textit{short} and \textit{long} method pairs.
We can see that the performance of all the approaches is worse on the long method pairs compared to that on the short method pairs.
This finding is consistent with the results of Tufano et al.~\cite{tufano2019empirical}.
Tufano et al.~\cite{tufano2019empirical} found that NMT-based APR can perform better on the shorter methods.
Note that our methods are longer than the dataset provided by Tufano et al..
The results from our experiments and Tufano et al. both indicate that the longer the method, the more complex the program logic is. Hence, more difficult it is to produce the correct repair.

Comparing the results of the approaches on the short method pairs with their results on the whole test set, we can see that the BLEU, CodeBLEU, and accuracy achieved by all the approaches have increased.
Similarly, we find that the performance of all the approaches on the long method pairs is worse than that of the whole test set.
It suggests that localizing faults in longer methods is more challenging, and it is possible longer methods have potentially more faulty locations, where more transformations are needed to generate the correct repair.

\subsection{Lessons Learnt}
We identify several insights that we hope can inspire the development of specialized techniques for repairing API misuse.
Specifically, we consider the unique challenges faced by learning-aided APR techniques in repairing API misuse.

\textbf{Repairing API misuses in \textit{long} methods is challenging.}
Based on the result in Section~\ref{sec:failure}, we find that current PLMs lack the ability to repair long methods.
PLMs, such as CodeBERT, can only handle input less or equal to 512 tokens by default.
However, some \textit{long} methods have more than 512 tokens.
A simple truncation does not work well, as shown in Table~\ref{tab:length-complete}.
Moreover, filtering out long methods is not solving the problem.
It may also be unrealistic to completely disallow developers from writing long methods.
On the other hand, splitting long methods into chunks or focusing on lower granularity may help in repairing API misuses.

\textbf{Integrating project-specific information can potentially improve repair performance.}
Like Example 4 shown in Figure~\ref{fig:rq1-failure}, this type of bug is unrelated to the syntactic change, while is more related to the project information.
This is a unique challenge in learning-aided APR techniques.
Learning-aided techniques usually have a limitation on the input length.
It is infeasible to treat the whole project information as input.
More ways to integrate project information should be a future direction to explore.

\textbf{Prioritize frequent and non-trivial API misuses.}
As shown in Section~\ref{sec:result}, it is challenging to repair API misuses.
Considering various types of API misuses, it may not be possible to completely repair all of them.
Some types of API misuse can be considered trivial for developers yet not easy for machines to repair.
We believe it is not worth resorting to an automatic approach to repair all API misuses. 
Instead, automation efforts should be focused on frequently occurring API misuses. Some frequently occurring API misuses are common program errors. 
It would save a lot of manual work if an automatic approach could be integrated when developers are coding.
Other than frequently occurring API misuses, more emphasis should be put on non-trivial API misuses, such as those that have a high impact on security or require more expertise to repair.
As the next step, researchers can focus on a certain type of critical API misuse and develop a specialized tool for it.

Except for the three points mentioned above, adopting learning-aided APR techniques for API misuse also shares common challenges with test-suite-based APR techniques as presented by Kechagia et al. ~\cite{kechagia2021evaluating}.
Here, we name a few: (1) incorrect fault localization: without correctly locating the API misuse bugs, it is impossible to repair the misuse;
(2) multiple faulty locations: multi-locations bugs are challenging to repair.
\subsection{Threats to Validity}
\label{sec:threats}

\textit{Threats to internal validity} relate to the correctness of our experiments.
For SequenceR and Recoder, we directly used the default settings as the original paper unless specified otherwise.
For general-purpose PLMs, we implement the models based on the replication package released by Zeng et al.~\cite{zeng2022extensive}.
We believe the threats are minimal.

\textit{Threats to construct validity} relate to experimental bias.
Following the prior works~\cite{codegpt,zeng2022extensive}, we not only use the exact match and BLEU but also include the newly introduced CodeBLEU score, which is more suitable for code as the evaluation metrics.
While we admit that a more rigorous way of evaluating the generated methods is to actually compile them, it is not feasible in practice to evaluate a large number of methods.

\textit{Threats to external validity} relate to whether our findings can be generalized to other datasets or PLMs.
In this work, we experiment with bug fixes from Java projects.
The results may differ when we experiment with Python projects.
However, since the pre-trained models are not specifically designed for a certain PL, we believe the threat is minimal. 
In the future, we plan to conduct experiments on dataset containing other PLs.
Another threat to external validity is on the approaches' selection.
We select two state-of-the-art APR techniques: SequenceR and Recoder. 
They represent two types of architecture, i.e., Seq2Seq-based and tree-based model.
We experiment with 9 PLMs, and larger PLMs have been released recently (e.g., CodeT5 has released a larger version~\footnote{\url{https://huggingface.co/Salesforce/codet5-large}}).
In the future, we plan to experiment with large PLMs.

\section{Related Work}
\label{sec:related}

\subsection{Pre-trained Language Models for Program Repair}

Given the success of PLMs in NLP tasks, researchers in software engineering are exploring their potential for use in general APR. 
Several studies have proposed novel APR methods built upon PLMs~\cite{cure}, while others investigate alternative ways to leverage them~\cite{xia2022less}.
To improve APR, existing approaches often utilize additional information beyond the buggy code. 
For example, TFix~\cite{berabi2021tfix} is based on the T5 model~\cite{raffel2020exploring} and requires the error messages from error detectors such as ESLint as input to generate bug fixes. 
Additionally, specialized PLMs designed to boost code review progress have been proposed~\cite{codereviewer,coditt5}, but these PLMs are meant to solve a different type of APR that requires NL reviews. 
Instead of traditional fine-tuning, recent works aim to leverage the pre-training objective (MLM) in PLMs and close the gap between pre-training and repair~\cite{xia2022less,yuan2022circle}. 
Xia and Zhang~\cite{xia2022less} explore zero-shot learning for generating bug fixes by treating program repair as a cloze-style task, where the buggy line is masked and the PLM is required to fill in the missing code. 
They have explored various templates to mask lines.
Similarly, CIRCLE~\cite{yuan2022circle} uses a prompt-based template to convert APR into a "fill-in-the-blank" task.
In our work, we conduct a comprehensive evaluation on various PLMs with traditional fine-tuning. In the future, we aim to investigate alternative ways of applying PLMs to repair API misuse.

\subsection{Domain-specific Program Repair}
Other than general APR techniques considered in our work, several more studies have focused on repairing domain-specific program errors, such as concurrency errors~\cite{li2019dfix}, the web~\cite{mahajan2018automated}, security vulnerabilities~\cite{huang2019using}, Android applications~\cite{tan2018repairing} and regression bugs~\cite{tan2015relifix}.

Li et al. proposed a tool named DFix~\cite{li2019dfix}, which focuses on fixing timing bugs in distributed systems.
Fixing distributed timing bugs has unique challenges: (1) it cannot rely on traditional synchronization primitives, such as locks and conditional variables; (2) it requires global code changes.
DFix can automatically process distributed timing bug reports, analyzes the buggy system and generates patches through static program analysis.
At a high level, DFix systematically generates patches that handle observed buggy timing through rollbacks or fast-forwards.
At a low level, DFix uses static analysis to automatically decide where and how to observe buggy timing and where and how to conduct rollback or fast-forward.

Domain-specific program repair techniques usually need to consider the uniqueness of the domain.
For instance, Mahajan et al.~\cite{mahajan2018automated} proposed a new approach that can automatically generate CSS patches that improves the mobile-friendliness of a web page.
Existing approaches can detect automatically detect mobile-friendly problems, but they are unable to repair the problem.
It remains a manual effort to repair the mobile-friendly problems in a web page.
To address this issue, Mahajan et al. propose an approach that first builds graph-based models of the layout of a web page.
The constraints encoded by these graphs are used to find patches that can improve mobile friendliness while minimizing layout disruption.
The approach leverages unique aspects of the problem domain to quantify metrics related to layout distortion and parallelize the computation of the solution to identify the best patch efficiently.

Different from the prior mentioned works, our work focuses on API-misuse repair, which belongs to another branch of domain-specific program repair.

\subsection{Empirical Studies on Program Repair}
In recent years, several studies have empirically evaluated the effectiveness of APR techniques~\cite{durieux2019empirical,ding2020patching,motwani2020quality}.
Durieux et al.~\cite{durieux2019empirical} conducted a large-scale experiment that evaluates 11 Java test-suite-based APR techniques on 2,141 bugs from 5 benchmarks.
They found that these techniques can generate patches for a diverse number of bugs, and they are complementary to each other.
Moreover, they found that the APR techniques perform significantly better on Defects4J than on the other benchmarks.
Furthermore, they identified six primary reasons that these APR techniques fail to generate patches for bugs, including incorrect fault localization and multiple fault localization.

Motwani et al.~\cite{motwani2020quality} analyzed the effectiveness of APR techniques in real-world Java programs, especially on the defects made by the developers during their regular development process.
Some of their findings are (1) APR techniques do sometimes produce patches, while those patches often break untested or undertested functionality; (2) The produced patches often overfit to the provided test suite.
Their work outlines the shortcomings of existing APR techniques when applied in the real world.

Different from the empirical studies mentioned above, our work evaluates APR techniques in repairing a specific type of error, i.e., API misuse.

\section{Conclusion and future work}
\label{sec:conclusion}
In this work, we present an empirical study that evaluates 11 learning-aided APR techniques on their capability to repair API misuse.
We build a large-scale API misuse benchmark, which consists of two variants: the \textit{complete data} with 118,490 pairs of the buggy and fixed methods, and the \singledata with 54,510 pairs of the single-line buggy and fixed methods. 
We conclude several findings based on the empirical results of existing approaches.
We find that decoder-based and encode-decoder-based PLMs are more effective than encoder-based PLMs.
Among all the 9 PLMs we investigate, CodeT5 achieves the highest scores in terms of EM.
PLMs are more effective than the studied APR techniques.

We recommend that future work addresses the limitation of the current APR techniques by carefully considering handling long methods, improving bug localization, and prioritizing frequent and non-trivial API misuses.
In the future, we are also interested in investigating alternative ways to adopt PLMs for repairing API misuse, such as prompt tuning~\cite{wang2022no}.
By analyzing the results of the current approaches, we plan to propose new approaches to repair API misuse.
Our dataset and the code are publically available at \url{https://anonymous.4open.science/r/TOSEM-API-Misuse}.

\balance
\bibliographystyle{acm}
\bibliography{main}

\end{document}